\documentclass[prb,twocolumn,showpacs,preprintnumbers,amsmath,amssymb]{revtex4}
\usepackage{graphicx}% Include figure files
\usepackage{dcolumn}% Align table columns on decimal point
\usepackage{bm}% bold math

\begin{document}

\title{Spin wave resonances in antiferromagnets}

\author{H.\,-A.\,Krug von Nidda$^1$, L.\,E.\,Svistov$^2$, L.\,A.\,Prozorova$^2$}
\affiliation{Center for Electronic Correlations and Magnetism EKM,
Experimentalphysik V, Universit\"{a}t Augsburg, D--86135 Augsburg, Germany$^1$,
P.\,L.\,Kapitza Institute for Physical Problems RAS, 119334 Moscow, Russia$^2$}

\date{\today}

\begin{abstract}
Spin wave resonances with enormously large wave numbers
corresponding to wave vectors $k\sim10^5\div10^6$ cm$^{-1}$ are
observed in thin plates of FeBO$_3$. The study of spin wave
resonances allows one to obtain information about the spin wave
spectrum. The temperature dependence of a non-uniform exchange
constant is determined for FeBO$_3$. Considerable softening of the
magnon spectrum resulting from the interaction of magnons, is
observed at temperatures above 1/3 of the N\'{e}el temperature. The
excitation level of spin wave resonances is found to depend
significantly on the inhomogeneous elastic distortions artificially
created in the sample. A theoretical model to describe the observed
effects is proposed.
\end{abstract}


\maketitle

\section{Introduction}

The linear excitation of spin waves in an ideal infinite crystal is
possible only if the frequency of pumping field $\omega_p$ matches
the spin-wave frequency $\omega_k$ (the law of conservation of
energy) and their wave vectors are equal (the law of conservation of
quasi-momentum). Since the wave vector of a microwave field is
relatively small ($\leq 10^2$ cm$^{-1}$) the spin waves with large
wave vectors cannot be excited.

The presence of defects in a sample allows the law of conservation
of quasi-momentum to be broken and makes the linear excitation
possible. One kind of possible natural defects in magnets is given
by the sample edges. If the magnon free path is comparable to the
sample size, the magnetic excitation spectrum becomes discrete,
giving rise to a finite coupling of the uniform microwave field with
spin-wave modes of the sample with non-zero wave numbers.

For a plane-parallel plate, the microwave field can be coupled to
the modes with $k_{\parallel}=0$ and $k_{\perp}\neq 0$, {\it i.e.}
with oscillations of magnetization which are uniform in-plane and
non-uniform across the plate. The excitation of such oscillations
(standing spin waves) by a uniform microwave magnetic field
(referred to as a spin-wave resonance) was predicted by Kittel
\cite{Kitt} and proven experimentally.\cite{Seav} The
resonance-field values are determined not only by the bulk
properties of a magnet but also by the fixation of the magnetic
moments on its surface. The boundary conditions for spins fully
fixed on a crystal surface are determined by the following relation:
\begin{equation}
 k_{zn}=n\pi/d.
\end{equation}
Only modes with odd number of half-periods are coupled to the
microwave magnetic field, with the efficiency of this coupling
decreasing inversely proportional to the wave vector $k_{z}$. The
study of spin-wave resonances in ferro- and ferrimagnetic films
allows one to obtain information about non-uniform exchange
constants and surface properties of magnets.\cite{Gure}

This work describes the experimental investigation of the linear
excitation of spin-wave resonances with large wave numbers $k\sim
10^5 \div 10^6$~cm$^{-1}$ in a single crystal of the easy-plane
antiferromagnet FeBO$_3$.

The static and dynamic properties of the rhombohedral
antiferromagnet FeBO$_3$ (symmetry group $D_{3d}^6$, N\'{e}el
temperature $T_N=348$~K) with an easy-plane anisotropy are studied
in detail.\cite{Koty} The low frequency branch of the spin-wave
spectrum at static magnetic field $H\perp C_3$ is given by the
relation:
\begin{eqnarray}
\omega_{1,k}^2=\gamma^2
[H(H+H_D)+H_{\Delta}^2+\alpha_{\perp}^2k_{\perp}^2+\alpha_{\parallel}^2
k_{\parallel}^2],
\end{eqnarray}
where $\gamma$ is the gyromagnetic ratio, $H_D$ is the Dzyaloshinsky field,
$H_{\Delta}$ is a parameter determined by magnetoelastic coupling, $H$ is the
external field applied in the basal plane of the crystal, $\alpha_{\parallel}$
and $\alpha_{\perp}$ are non-uniform exchange constants, $k_{\parallel}$ and
$k_{\perp}$ are wave-vector components along the $C_3$-axis and in the basal
plane respectively. The values of these constants at $T=77$~K are the
following: $\gamma =2\pi\cdot 2.8$~GHz/kOe, $H_D\simeq 100$~kOe,
$H_{\Delta}\simeq 1.9$~kOe, $\alpha_{\parallel}=7.8\cdot
10^{-2}$~Oe$\cdot$cm.\cite{Koty}

The observability of parametric excitation of spin waves shows that
the lifetime of magnons with the frequency $\omega_k/2\pi\simeq
10^{10}$~Hz and wave vector $\mathbf{k}= 0\div 10^6$~cm$^{-1}$ is
rather long at helium temperatures: $\tau\sim 0.1\div 1$~$\mu$sec.
The velocity of magnons with $\mathbf{k}= 10^5-10^6$~cm$^{-1}$ is
$s_k=\mid\partial\omega_k/\partial k\mid\sim 10^5-10^6$~cm/sec.
Using these values one can estimate the magnon free path
$\lambda=s\tau\simeq 1$~mm. Therefore, samples with the thickness
smaller than 0.1~mm should be used to observe spin-wave resonances.

\section{Samples and experimental technique}

Single crystal plates with the size of approximately $0.02\times
3\times 4$~mm$^3$ were used in our experiments. Their developed
faces coincided with the basal plane and were optically smooth. The
sample quality was checked by x-ray topography. Only homogeneous
single-domain samples were chosen for the measurements.

The investigations were carried out using a standard Q-band Bruker
spectrometer. The field derivative of the transmitted signal $dP/dH$
was measured as a function of external magnetic field $H$. The
measurements were performed in the temperature range 4.2 to 280~K.

\section{Spin wave resonances in undistorted FeBO$_3$ samples}

\begin{figure}
\begin{center}
\includegraphics[width=\columnwidth]{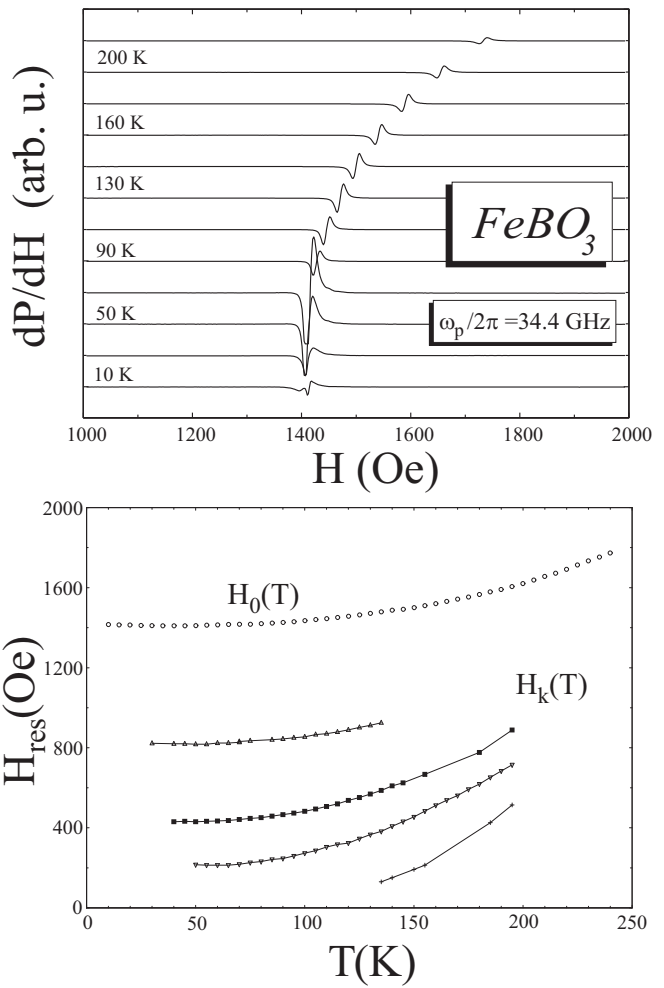}
\caption{Upper panel: the field dependence of derivatives of the
transmitted signal $dP/dH$ recorded at various temperatures. Lower
panel: the temperature dependences of an antiferromagnetic resonance
and four spin-wave resonance fields traced in a wide temperature
range (corresponding records are shown in Fig.~\protect\ref{fig:2});
the absorption lines at various temperatures are rescaled and
shifted.} \label{fig:1}
\end{center}
\end{figure}

The experiment was done on a single-crystal plate of $d=0.02$~mm in
thickness at a frequency $\omega_p/2\pi =34.4$~GHz. Since the
resonance absorption in FeBO$_3$ was found to depend on the way of
sample mounting, we used the most delicate one. The thin-walled
glass tube was filled by chemically pure fine powder of sodium
chloride. The sample was put horizontally and covered by salt to
prevent it from motion. The tube was inserted into the resonator.
This mounting procedure minimizes the sample deformation which can
be checked by the antiferromagnetic resonance (AFMR) linewidth. The
upper panel of Fig.~\ref{fig:1} shows the AFMR line records taken at
various temperatures. The external magnetic field $\mathbf{H}$ and
the microwave field $\mathbf{h}$ are mutually perpendicular and lie
in the basal plane of the crystal. On increasing the temperature,
the line shifts to larger fields. The temperature dependence of the
resonance field $H_0$ is represented on the lower panel of
Fig.~\ref{fig:1}. The AFMR linewidth is $\Delta H =12\pm 1$~Oe in
the whole temperature range. At $T<30$~K the line becomes
asymmetric.

\begin{figure}
\includegraphics[width=\columnwidth]{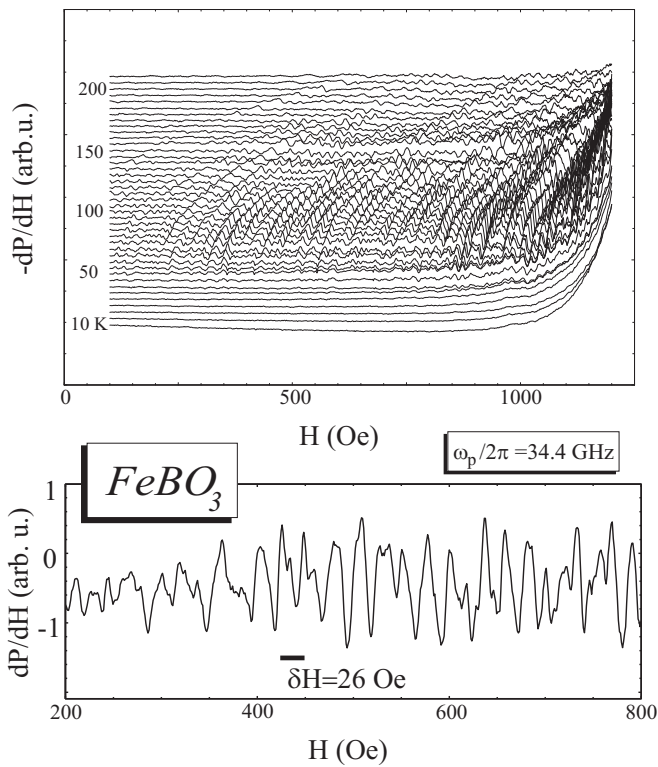}
\caption{The field records of derivatives of the transmitted signal
at various temperatures.} \label{fig:2}
\end{figure}

The field dependences of derivatives of the transmitted signal
$dP/dH$ taken at various temperatures are presented in
Fig.~\ref{fig:2}. Regular absorption lines are observed on these
records, with their positions shifting to larger fields on heating,
analogously to the AFMR lines. The sensitivity of the spectrometer
does not depend on the temperature of the record. Several most
pronounced lines were traced in a wide temperature range. The
corresponding dependences of four resonance fields are shown on the
lower panel of Fig.~\ref{fig:1}. Spin-wave resonances with large
wave vectors were observed up to temperatures $\sim 250$~K. The
spin-wave resonances start to decrease in amplitude below $T\sim
50$~K and become practically unresolvable at $T\lesssim 30$~K. The
linewidth of a single spin-wave resonance is $\approx 6$~Oe in the
vicinity of $H_0$ and $\approx 10$~Oe in low fields.

The fragment of the record $dP/dH(H)$ at $T=100$~K is shown on the
lower panel of Fig.~\ref{fig:2}. The distance between neighboring
spin-wave resonance fields calculated by formula~(2) is marked by a
segment. One can see that spin-wave resonances with both odd and
even wave numbers are excited.

\section{Spin wave resonances in FeBO$_3$ samples with inhomogeneous distortions}

\begin{figure}
\includegraphics[width=\columnwidth]{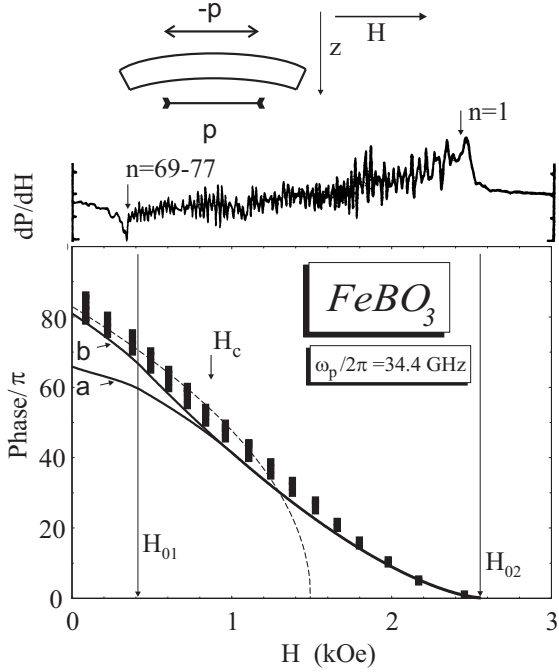}
\caption{Upper panel: Calculated field dependences
$n(H)=Phase/\pi=\int^{d}_{0}k(z)dz / \pi$ (solid lines~a,b).
Spin-wave resonances are expected near fields at which the phase is
divisible by $\pi$. The curve~`a' is calculated taking into account
the rotation of an antiferromagnetic vector at fields $H<H_{c}$ and
under the condition $\mathbf{p}\parallel \mathbf{H}$; the curve~`b'
is calculated under the condition $\mathbf{l}\perp\mathbf{H}$ in the
whole sample at all fields $H$. The calculated $n(H)$ dependence for
an undistorted sample is shown by the dashed line. Black rectangles
correspond to the $n(H)$ dependence obtained from the experimental
curve shown on the upper panel of the Figure; $T=80$~K,
$d=0.013$~mm, $\omega_p/2\pi =34.4$~GHz.} \label{fig:4}
\end{figure}

Our samples were thin plates with their thickness being at least 100
times smaller than the other dimensions. Application of a glue onto
one of the developed sample planes creates a tension which is
non-uniform across the plate: the sample is compressed from one side
and stretched from the other (see insert in Fig.~\ref{fig:4}). Since
the thermal expansion coefficients of the glue and the sample are
different, the value of the tension depends on temperature. Narrow
resonance lines are observed on the background of a broadened AFMR
line, becoming denser and more intense in the vicinity of the
resonance field $H_0$. This fine structure is especially pronounced
around 100~K and disappears below 30~K.

\begin{figure}
\includegraphics[width=\columnwidth]{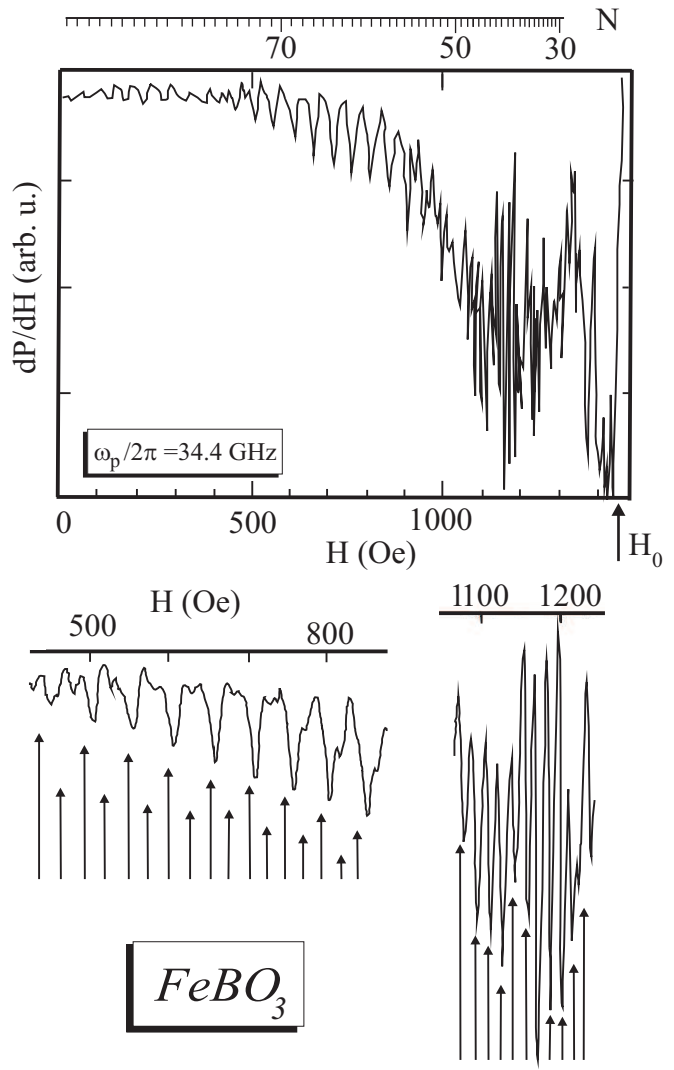}
\caption{Fragment of the derivative of the transmitted signal
$dP/dH$ measured at $T=80$~K in the sample with one of the sides
covered by a thin film of glue. The positions of spin-wave
resonances with the wave number $k_z=\pi n/d$ ($d=0.016$~mm is the
plate thickness) obtained by formulae (1,2) are given on the upper
scale. The lower panel shows two expanded fragments of this record.
The calculated resonance field values are marked by arrows.}
\label{fig:3}
\end{figure}

Fig.\ref{fig:3} shows a fragment of the AFMR line recorded at
$T=80$~K. The positions of the spin-wave resonances calculated by
formula~(2) with the above values of constants and $k_z=\pi n/d$
(the plate thickness $d=0.016$~mm) are given on the upper scale of
this Figure. The field intervals between the neighboring resonance
features are in good agreement with these calculations. The
spin-wave resonances observed in the low-field range correspond to
numbers $n\simeq 80$, {\it i.e.} $k_z\simeq 1.5\cdot
10^5$~cm$^{-1}$. The spin-wave resonances with $n\leq 20-30$ in the
vicinity of $H_0$ are not resolved. The modes with odd and even
number of half-periods $n$ are excited in the vicinity of an AFMR
field with roughly the same efficiency, while apart from $H_0$ each
other resonance was considerably weaker. The intensity of the
spin-wave resonances in a sample with inhomogeneous distortion is at
least by two orders of magnitude larger than that in an undistorted
sample.

The influence of uniaxial stress on the spin-wave spectrum in
easy-plane antiferromagnets was studied both experimentally and
theoretically.\cite{Boro5, Turo2, Turo3} It was shown that the
effect of uniaxial stress $\mathbf p$ in the basal plane of the
crystal can be described by an effective magnetic field $\mathbf
H_{me}(\mathbf p)$. The additional gap $H_{\Delta 1}^2$ arising in
the spin-wave spectrum is related to this field by the following
expression:
\begin{equation}
H_{\Delta 1}^2 =2H_E H_{me}(p),
\end{equation}
where $H_E$ is the exchange field. Thus, even weak distortions can
significantly modify the spin wave spectrum in such antiferromagnets
due to ``exchange amplification''. In case of a uniaxial distortion,
the magnetoelastic gap changes across the plate, so that the wave
vector of a spin wave propagating perpendicular to the basal plane
should depend on $z$-coordinate.

Let us discuss how the spin-wave resonances change in the presence
of such distortions. The spin-wave resonance condition can be
expressed as follows:\cite{Gure}
\begin{equation}
\int^{d}_{0}k(z,p,H)dz=\pi n,\label{SWRES}
\end{equation}
where $n$ is an integer determining the number of the spin-wave
resonance. This equation is written under the assumption that the
spins are fully fixed at sample edges. Supposing that the value of a
uniaxial stress varies linearly from $-p$ to $p$, one can calculate
the spin-wave resonance fields for a given plate thickness and
compare them to the experiment. The value of an uncontrollable
parameter $p$ can be estimated by the position of the features
observed on the field dependences of $dP/dH$ at $H=H_{01}$ and
$H=H_{02}$ (see the experimental curve on the upper panel of
Fig.~\ref{fig:4}). We relate these features to fields at which the
condition $\omega _p\simeq\omega(k\simeq 0,H)$ is satisfied in the
vicinity of upper and lower edges of the crystal. The calculated
field dependences of $Phase/\pi=\int^{d}_{0}k(z)dz / \pi$ are shown
on the lower panel of Fig.~\ref{fig:4} by solid lines~a,b. The
spin-wave resonances are expected around the field at which the
phase is divisible by $\pi$. The experimentally obtained spin-wave
resonance numbers $n$ at some fields $H$ are shown on the same
Figure by black rectangles. The height of the rectangle corresponds
to the error in experimental determination of $n$. The experimental
$n(H)$ dependence is well described by the proposed model. In the
low field range the function $n(H)$ strongly depends on the angle
between the vectors $\mathbf{p}$ and $\mathbf{H}$. The observed
discrepancy between the experiment and model curve~`a' at low fields
is possibly associated with unparallel orientation of these vectors.

\begin{figure}
\includegraphics[width=\columnwidth]{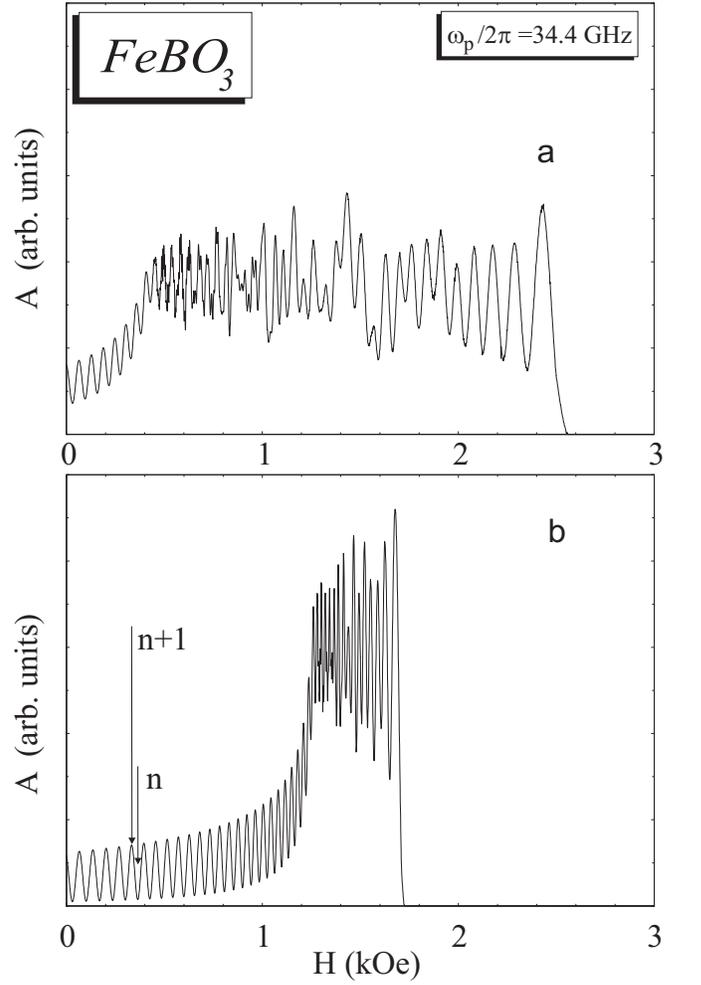}
\caption{Calculated field dependencies of a coupling coefficient
of spin wave resonances with a uniform microwave field for two
values of non-uniform distortions. Panel `a' corresponds to the
same value of $p$ as in Fig.~\protect\ref{fig:4} at $H_c=200$~Oe,
the value of $p$ on panel `b' corresponds to
Fig.~\protect\ref{fig:3} at $H_c=100$~Oe.} \label{fig:5}
\end{figure}

Fig.~\ref{fig:5} shows the calculated field dependencies of a
coupling coefficient $A(H)$ of spin wave resonances with a uniform
microwave field for two values of non-uniform distortions. Panel
`a' shows the calculation with the same parameter $p$ as in
Fig.~\protect\ref{fig:4}, while that on panel `b' is the same as
in Fig.~\protect\ref{fig:3}. This coefficient is calculated as
follows:
\begin{equation}
A(p,H)=A_0\int^{d}_{0}\sin (\int^{z}_{0}k(z,p,H)dz)dz
.\label{Ampl}
\end{equation}
One can see that the spin wave resonances with both odd and even
wave numbers are coupled to the microwave field due to non-uniform
distortion of the sample. The amplitudes of spin wave resonances
calculated in the frame of the discussed model are irregular at
fields $H_{c1}<H<H_{c2}$ and regular in the low field range. Thus,
the above model is in satisfactory agreement with experimentally
observed spectra.

\section{Conclusion}

\begin{figure}
\includegraphics[width=\columnwidth]{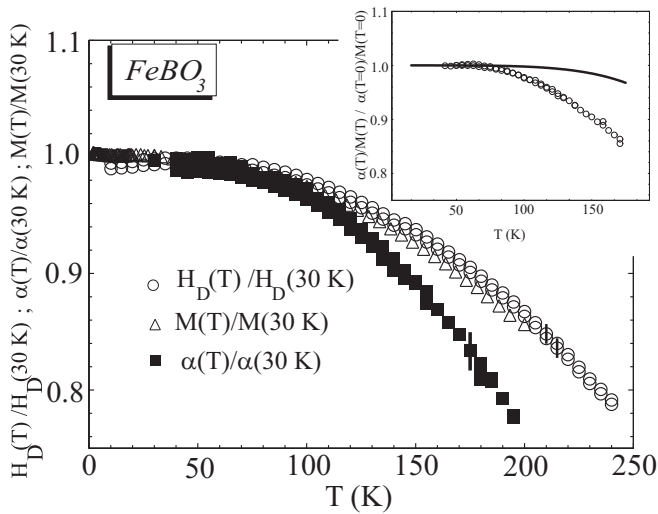}
\caption{Temperature dependencies of a Dzyaloshinsky field $H_D(T)$
({\Large $\circ$}) determined from AFMR measurements, a non-uniform
exchange constant $\alpha (T)$ ($\blacksquare$) and magnetic moment
of FeBO$_3$ sample in a field $H=1.5$~kOe applied in the basal plane
of the crystal ($\triangle$). All dependencies are rescaled to
values of corresponding parameters measured at $T=30$~K. The inset
shows the correction to the non-uniforms-exchange constant
$\alpha/M$ due to interaction of magnons calculated in
\protect\cite{Sobd}, and the same ratio obtained from the
experimental results shown on the Figure.} \label{fig:6}
\end{figure}

In conclusion, the spin wave resonances with $k\approx 1.5 \cdot
10^5$~cm$^{-1}$ are observed in single-crystal plates of FeBO$_3$.
The fact, that the spin-wave resonances observed in samples with
non-uniform distortions are at least 100 times more intensive than
those in undistorted ones demonstrates the influence of an elastic
tension on the coupling of a standing spin wave with microwave
field. Besides, the non-uniform distortion inevitably leads to the
coupling between the microwave pumping and resonances both with odd
and even wave numbers $n$ (see Fig.~\ref{fig:5}). The spin wave
resonances with even $n$ were also observed in undistorted samples
which probably results from internal tensions in these crystals.

Spin-wave resonances were clearly observed in the temperature
interval $30\div 250$~K. The resonance line can be recorded, if its
linewidth is smaller than the distance between neighboring lines.
The upper temperature limit is determined by enhanced damping of
spin waves due to three magnon processes.\cite{Koty} The lowest
boundary is stipulated by the proximity to the damping peak at
$T=18$~K resulting from the ``low relaxation'' process due to
Fe$^{2+}$ impurities in FeBO$_3$ crystals.\cite{Koty} Since the
spin-wave resonances are well resolved in a wide temperature range,
one can trace the temperature dependence of the resonance field for
resonances with large wave numbers. The corresponding data for
several wave numbers are given on the lower panel of
Fig.~\ref{fig:6}. Assuming the spin-wave spectrum in the whole
temperature interval to be described by formula~(2), one can obtain
the temperature dependence of the non-uniform exchange constant
$\alpha _{\parallel}(T)$.

Fig.~\ref{fig:6} demonstrates the temperature dependences $\alpha
_{\parallel}(T)$ and $H_D(T)$. As seen from this Figure, the
$H_D(T)$ dependence coincides within the experimental accuracy with
the dependence of a spontaneous magnetic moment $M(T)$ measured in
the same sample by standard SQUID-magnetometer. The value of the
non-uniform exchange constant $\alpha _{\parallel}$ at $T>100$~K
decreases considerably faster on increasing the temperature. The
effect of three- and four magnon interaction processes on the magnon
spectrum in an easy-plane antiferromagnet was studied
earlier.\cite{Sobd} The upper panel of Fig.~\ref{fig:6} shows the
correction to the exchange constant $\alpha/M$. The calculated
decrease of the exchange constant is about 5 times smaller than that
experimentally found at $T=150$~K. The observed discrepancy can be
associated with the assumption made in Ref.~\onlinecite{Sobd} that
the low magnon branch is gapless ($H=0$, $H_{\Delta}^2=0$).

The authors thank V.N. Seleznev for preparing single crystal samples of
FeBO$_3$. This work is supported by RFBR, grant 07-02-00725 and by
German Research Society (DFG) within the Transregional
Collaborative Research Center (TRR 80).

\end{document}